\documentclass[useAMS,usenatbib]{mn2e}
\usepackage{epsfig,amssymb}
\title[Gamma-ray halo of 3C~279]{
Constraining the extension of a possible gamma-ray halo of 3C~279 from
2008--2014 solar occultations} \author[E.~Kotelnikov, G.~Rubtsov and
S.~Troitsky]{
Egor Kotelnikov$^{1}$, Grigory Rubtsov$^{2}$ and
Sergey Troitsky$^{2}$\thanks{E-mail: st@ms2.inr.ac.ru}\\
$^{1}$Physics Department, M.V.~Lomonosov Moscow State University,
Vorobjevy Gory, 119991, Moscow, Russia\\
$^{2}$Institute for Nuclear Research of the Russian Academy of
Sciences,
60th October Anniversary prospect 7a, 117312, Moscow, Russia
}
\begin{document}
\date{2015 March 12, in original form 2014 December 03}
\pagerange{L\pageref{firstpage}--L\pageref{lastpage}} \pubyear{2015}
\maketitle
\label{firstpage}
\begin{abstract}
The angular extension of the gamma-ray image of 3C~279 may be constrained
by studying its solar occultations as suggested by \citet{FRT-MN}. We
perform this kind of analysis for seven occultations observed by Fermi-LAT
in 2008--2014, using the Fermi-LAT Solar System tools. The results are
interpreted in terms of models with extended gamma-ray halo of 3C~279;
first constraints on the size and the flux of the halo are reported.
\end{abstract}


\section{Introduction}
\label{sec:intro}
Some time ago, it was pointed out \citep{FRT-PRL} that one of the
brightest gamma-ray sources in the sky, blazar 3C~279, is screened by the
Sun every year. Since the Sun is not a very bright source of photons with
energies $\gtrsim 100$~MeV, these events look as occultations and may be
used to constrain some new models of physics and astrophysics. In
particular, an extended halo of the source, if exists, would shine beyond
the solar disk while the central point source is screened. Measurements of
the gamma-ray flux of 3C~279 during the occultation may constrain
\citep{FRT-MN} the size and the flux of the extended halo in a way similar
to the early measurements of angular diameters of stars in their lunar
occultations, with the precision exceeding the angular resolution of the
instrument.

Recently, the Fermi LAT collaboration has performed \citep{Fermi-occ} an
analysis of four (2008--2011) solar occultations of 3C~279 and constrained
the source flux during the periods when its center was screened.
\citet{Fermi-occ} have considered and constrained two scenarios:
(i)~transparency of the Sun for gamma rays and (ii)~an extended
sharp-edged disk with constant surface brightness and a priori fixed
energy-dependent radius of $1.5^{\circ} \times \left(E/500~{\rm MeV}
\right)^{-1}$, where $E$ is the gamma-ray energy.
Nevertheless, there is a gap between these interesting results and
constraints on realistic models of physics and astrophysics. We note that
the solar transparency, case (i), as suggested by \citet{FRT-PRL},
requires an axion-like particle with parameters firmly excluded, since
then, by several laboratory experiments, see e.g.\ \citet{PVLAS-excl},
\citet{OSQAR-excl} and \citet{ALPS-excl}. The extended halo, case (ii), if
it exists, should be formed as a result of a random scattering process and
is unlikely to look like a sharp-edged constant-intensity disk. The size
of the halo is determined by a number of physical parameters of the source
and/or of the intergalactic space, notably by the values of the magnetic
fields. The halo extension is therefore an important observable to be
constrained while it was kept a priory fixed in the analysis of
\citet{Fermi-occ}.

In the present work, we fill this gap and follow the prescription of
\citet{FRT-MN} more precisely; namely, we consider the halo size as a free
parameter, while the halo shape is taken to be Gaussian. This approach
opens a way to study and to constrain various models of the halo formation,
as well as physical parameters of the source and the strength of
the intergalactic magnetic field; note that the latter is a subject of
intense debates. We make use of publicly available Fermi-LAT data for
seven occultations (2008--2014) thus almost doubling the statistics as
compared to four occultations used by \citet{Fermi-occ}.

\section{Data, analysis and results}
\label{sec:anal}
In this work, we use the Pass 7REP (V15) data of Fermi LAT
\citep{Fermi-LAT}. We consider seven solar occultations of 3C~279 which
happened on the 8th of October, 2008--2014. The precise periods of the
occultations have been calculated with the help of the HORIZONS system
\citep{HORIZONS}. We processed the data with the standard \textit{gtlike}
routine from \textit{Fermi Science Tools v9r32p5}\footnote{%
{\tt http://fermi.gsfc.nasa.gov/ssc/data/access/}.}.
We use the ``SOURCE'' class photons with energies greater than 100~MeV. We
apply standard quality cuts and require that photon zenith angles with
respect to the Earth do not exceed $100^\circ$ and the satellite rocking
angle is smaller than $52^\circ$.

Since we are interested in solar occultations, we need to disentangle
fluxes of the Sun and of the blazar during the events. The Sun is a
gamma-ray source firmly detected by Fermi LAT \citep{Fermi-Sun2011}, with
the flux from the solar disk being produced by cosmic-ray interactions
with the solar surface \citep{disk1,disk2,disk3}. Another, much more
extended, flux component is caused by the inverse Compton scattering of
cosmic-ray electrons on the solar light \citep{Moskalenko,OS1,OS2}. At the
time scales of order a week to a month, the flux of the quiet Sun is
stable and may be determined by means of dedicated tools developed by the
Fermi group for observations of moving targets, the Fermi-LAT Solar System
Tools \citep{1307.0197}.

\begin{figure}
\begin{center}
\includegraphics[width=0.95 \columnwidth]{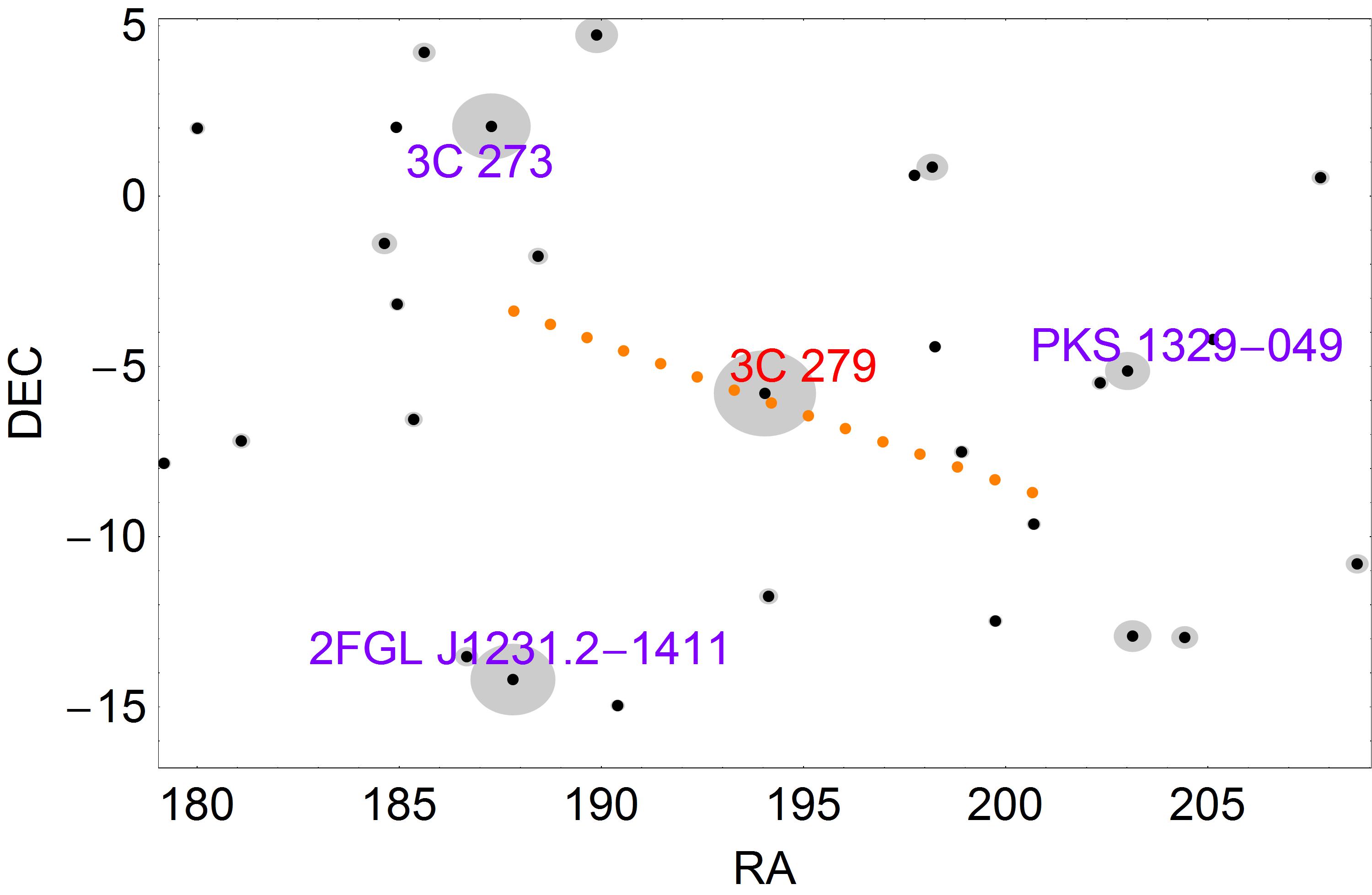}
\caption{
\label{fig:path}
A map of the gamma-ray sky around 3C~279. Black dots correspond to
gamma-ray sources listed in the 2FGL catalog \citep{2FGL}, the area of grey
circles around them is proportional to their 2FGL flux. The solar path
within $\pm$1 weeks from the occultaion is shown by orange dots.}
\end{center}
\end{figure}

However, the time scale of an occultation, $\sim 8.5$~hours, is too short
to determine its
flux with the required accuracy. We, therefore, choose to study a four-week
period surrounding each occcultation, to determine the solar flux by means
of the Solar System Tools.
The tools assume that the solar flux is proportional to an average Fermi
LAT solar flux. The spatial-spectral template of the moving Sun for the
given time interval is produced with the {\it gtsuntemp} tool. The
template allows to consider the Sun as an extended source in the {\it
gtlike} tool, with the normalization treated as a free parameter.
The template includes the solar disk as well as the extended emission with
their energy and direction dependencies, while a single parameter, the
overall flux normalization, is fitted. A complete model of the sky region
includes Sun, galactic and isotropic diffuse components, 3C~279 and other
point sources in the area of interest from the 2FGL catalog~\citep{2FGL}.
The solar path in the gamma-ray sky is sketched in Figure~\ref{fig:path}.
The results are given in Table~\ref{tab:fluxes},
\begin{table}
\centering
\begin{tabular}{|c|c|c|c|}
\hline
\hline
& \multicolumn{2}{|c|}{Flux, $\pm 2$ weeks,} & Flux at occultation,\\
Year & \multicolumn{2}{|c|}{$10^{-7}$ cm$^{-2}$ s$^{-1}$} &  $10^{-7}$
cm$^{-2}$ s$^{-1}$, \\
& Solar disk & 3C~279 & 3C~279\\
\hline
2008 & 4.6$\pm$0.4 & 3.6$\pm$0.4 & 2.5$\pm$2.6\\
2009 & 5.2$\pm$0.5 & 7.1$\pm$0.6 & 5.0$\pm$5.3\\
2010 & 3.0$\pm$0.4 & 8.1$\pm$0.6 & 1.7$\pm$4.6\\
2011 & 3.2$\pm$0.4 & 2.6$\pm$0.3 & 0.8$\pm$1.5\\
2012 & 4.9$\pm$0.4 & 2.4$\pm$0.4 & 2.1$\pm$2.7\\
2013 & 7.4$\pm$0.4 & 6.2$\pm$0.6 & 2.0$\pm$2.1\\
2014 & 3.9$\pm$0.4 & 3.8$\pm$0.4 & 5.5$\pm$3.1\\
\hline
stacked & 4.7$\pm$0.2 & 4.0$\pm$0.2 & 2.6$\pm$1.1\\
\hline
\end{tabular}
\caption{\label{tab:fluxes}
Fluxes ($E>100$~MeV) of the solar disk and of 3C~279, as determined for
$\pm 2$ weeks around the occultation, and of 3C~279 during the occultation;
see the text for details of the account of the solar background. The flux
of the solar extended emission, integrated over a large area up to
20$^{\circ}$ from the Sun, is 1.48 times the disk flux. The last row gives
the fluxes for the sum of 7 stacked exposures. }
\end{table}
together with 3C~279 fluxes for the same
periods, quoted for reference\footnote{One may note that the solar flux
quoted for 2013 was considerably higher than in other periods. We have
traced the reason for this. Large solar flares, of M class and higher, are
confirmed sources of energetic gamma rays, see e.g.\ \citet{flares}. The
enhancement of the solar flux in the 2013 period corresponds to a series
of solar flares between the 9th and the 14th of October, 2013,
recorded in the NOAA archive, {\tt http://www.solarmonitor.org}.
Both the detected flares and other unaccounted solar activity
may affect systematically the signal and background estimates. We
have checked that no flares were detected during all seven
occultations and the systematic effect of the 2013 solar flares is within
10\% in terms of the final result. We leave the investigation of a more
detailed time-dependent solar flux model for future.}. Then, we use
this fixed solar flux in the background model for the short-time
observations during the periods when the point source at the position of
3C~279 was screened by the Sun. The flux of 3C~279 during the
occultations, obtained in this way, is given in the last column of
Table~\ref{tab:fluxes}. Within the statistical errors, this flux is
consistent with zero for each individual occulatation, as one would expect
for a usual point-like source. In order to achieve better precision, we
merge the seven occultations with the following procedure:
\begin{enumerate}[a]
\item the 7 photon files corresponding to the particular occultations are
used jointly,
\item the exposure cubes are coadded with the {\it gtltsum} tool,
\item the 7 extended sources are included in the model, each representing
the solar flux and the motion template for a particular year.
\end{enumerate}
A stacked sum of all seven
exposures results in a marginal excess of $\left(2.6 \pm 1.1 \right)
\times 10^{-7}$~cm$^{-2}$s$^{-1}$, the 95\% confidence level (CL) limit on
the observed flux of 3C~279 when it is screened by the solar disk is thus
rather weak,
\begin{equation}
F_{\rm obs}<4.8 \times 10^{-7}~{\rm cm}^{-2}{\rm s}^{-1}~~(95\%~{\rm CL}).
\label{Eq:flux-limit}
\end{equation}
The best-fit excess in the stacked occultation result, according to the
\textit{gtlike} tool, corresponds to $\approx 22$ photons from the source
while the number of background photons from the Sun is $\approx 90$. The
stacked solar flux (both for the disk and for the extended emission),
obtained in our fits, is $(1.02 \pm 0.04)$ times the value of
\citet{Fermi-Sun2011}.

\begin{figure}
\begin{center}
\includegraphics[width=0.95 \columnwidth]{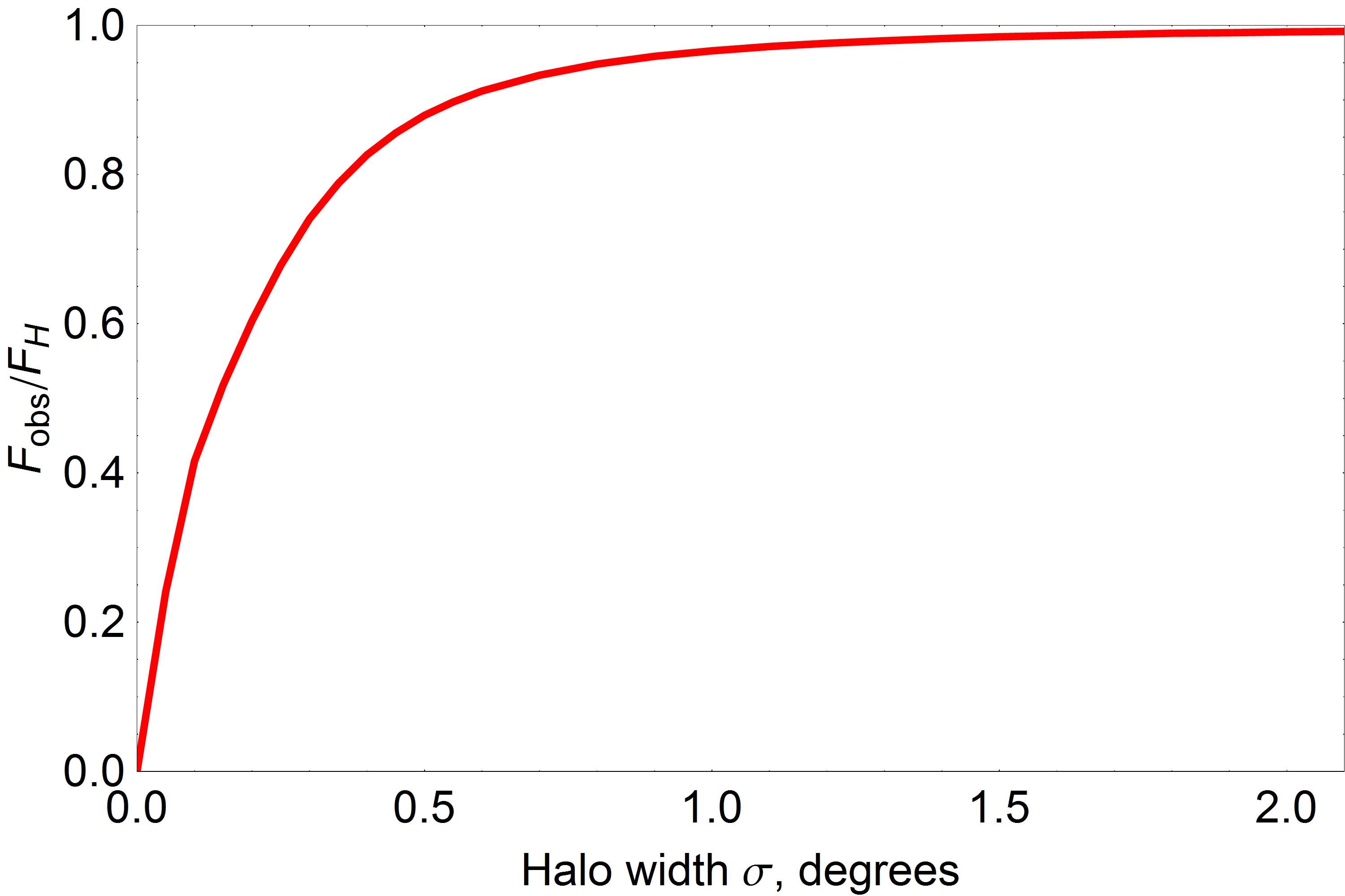}
\caption{
\label{fig:fraction}
Fraction $F_{\rm obs}/F_{\rm H}$ of the flux of a Gaussian extended source
centered at 3C~279 and not screened by the solar disk during the
occultation, as a function of the image extension $\sigma$. }
\end{center}
\end{figure}

\begin{figure}
\begin{center}
\includegraphics[width=0.95 \columnwidth]{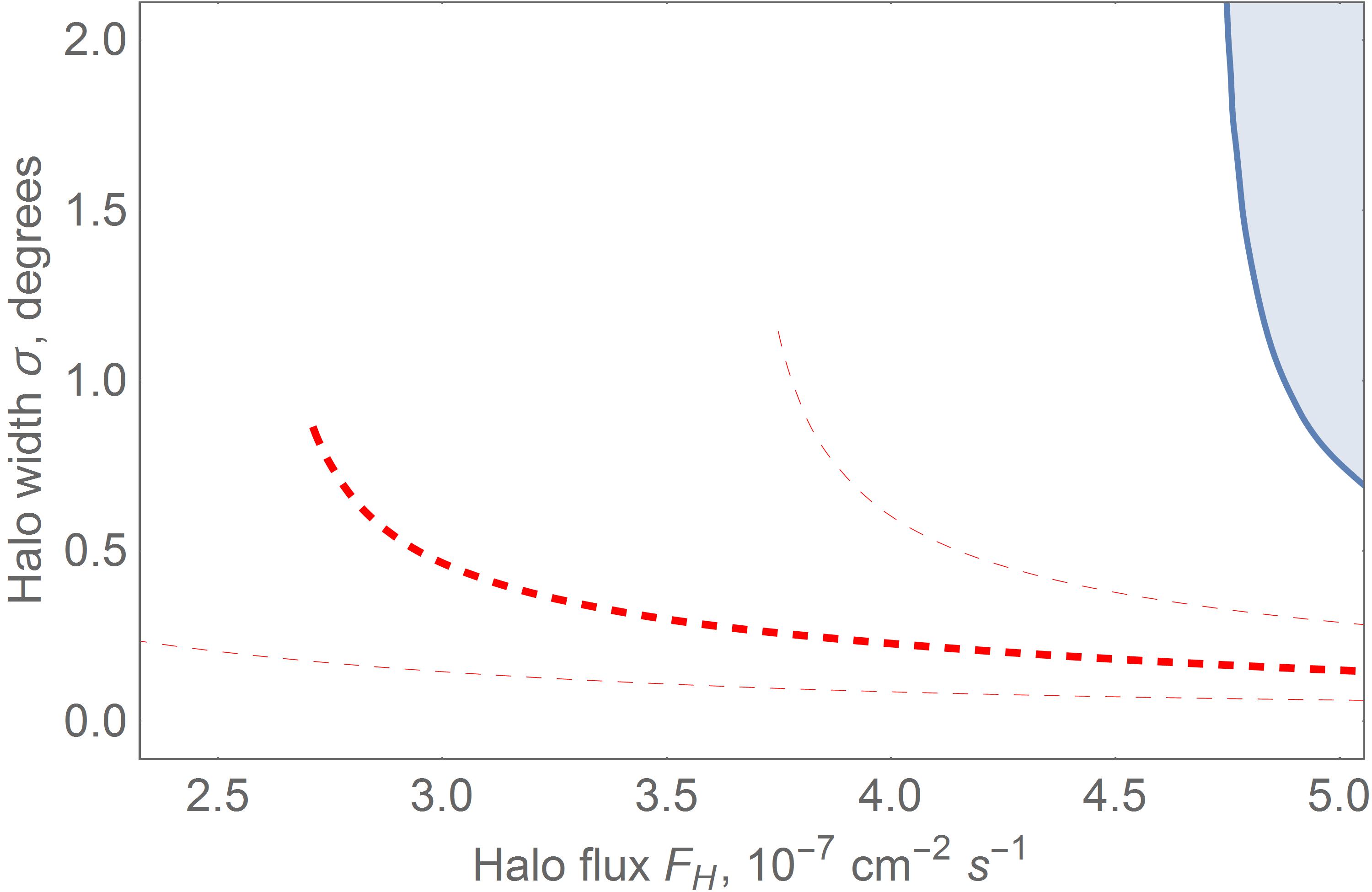}
\caption{
\label{fig:constraints}
Constraints on the extension of the Gaussian halo of $E>100$~MeV photons
around 3C~279, as obtained in this work. Shaded area is excluded at the
95\% confidence level; thick dashed line corresponds to the best-fit halo
extension; thin dashed lines limit the 68\% confidence-level range around
the best fit.}
\end{center}
\end{figure}

To interpret the flux constraints in terms of the extended image, we
simulate photons from a Gaussian extended source with the total flux
$F_{\rm H}$, centered at the position of 3C~279, with the extension
$\sigma$. The flux density at the angular distance $\theta$ from the
blazar is thus proportional to
$$\exp\left(-{\theta^2 \over 2\sigma^{2} }\right).$$
Then we take into account the apparent motion of the
Sun across the source and calculate the fraction of the simulated photons
which are not screened by the solar disk during the occultation, when the
central point is behind the Sun. This fraction, $F_{\rm obs}/F_{\rm H}$,
is plotted in Fig.~\ref{fig:fraction}
as a function of $\sigma$. By making use of this plot, we directly
constrain $F_{\rm H}$ and $\sigma$ from $F_{\rm obs}$, as prescribed by
\citet{FRT-MN}. The results are shown in Fig.~\ref{fig:constraints},
and these constraints on the extended image represent the main result of
our present study.

\section{Conclusions}
\label{sec:concl}
In this Letter, we report on the analysis of Fermi-LAT data for
seven solar occultations of 3C~279. Our results, expressed in terms of the
flux coming from the screened source,
Table~\ref{tab:fluxes}, are in agreement with earlier studies (four
occultations seen by Fermi LAT, \citet{Fermi-occ}, and one occultation
seen by EGRET, \citet{FRT-PRL}, reanalized with the account of the solar
flux by \citet{Fermi-occ}). However, we use these results, for the first
time, to obtain constraints on the size of a possible extended gamma-ray
halo of the source. These constraints are presented in the flux--extension
parameter space, Fig.~\ref{fig:constraints}. Implications of these results
for numerous particular models of the halo formation, see e.g.\
\citet{ACV-1994}, \citet{GA-2007}, \citet{AD-2008}, \citet{NS-2009} and
numerous more recent elaborations, will be discussed elsewhere. It is
interesting to note that, with the enhanced statistics we use (seven events
versus four or one), a marginal excess consistent with the extended image
is seen, though with a low statistical significance of $\sim 2.4$ standard
deviations. This excess, if it were real, would be consistent with a halo
of $\sim (0.5^{\circ}-1^{\circ})$ extension and $\sim (2-3)\times
10^{-7}$~cm$^{-2}$s$^{-1}$ total flux, as well as with the solar
transparency in the axion scenario (which predicts 1/3 of the photon beam
to shine through the Sun\footnote{Interestingly, the inverse-Compton
emission is essentially zero in the direction of the solar disk, thus
reducing the background in the searches for this effect
\citep{stellarics}. Just outside the disk, the emission from beyond the Sun
returns to a significant level, so it is necessary to account for this
background in the present study.} but for which a consistent description in
terms of a particle not excluded by laboratory experiments is missing). At
higher energies, $E>1$~GeV, the present statistics is too low to put an
interesting limit. Further studies of solar occultations of 3C~279,
notably in the pointing mode of Fermi LAT, and of the image extension of
the blazar in its low state are needed in order to confirm or falsify
these results.

\section*{Acknowledgments}
The authors are indebted to
O.~Kalashev
for interesting discussions.
This work was supported by the Russian Foundation for Basic Research, grant
13-02-01293 (adaptation of the method of solar occultations to the
Fermi-LAT data, E.K.), and by the Russian Science Foundation, grant
14-12-01340 (constraining the extension of the gamma-ray image of the
blazar 3C~279, G.R.\ and S.T.).  G.R.\ acknowledges the fellowship of the
Dynasty foundation. The analysis is based on data and software provided by
the Fermi Science Support Center. The numerical part of the work has been
performed at the cluster of the Theoretical Division of INR RAS.

\bsp

\label{lastpage}

\end{document}